# Antiproton- hydrogen collisions calculation by Coulomb wave function discrete variable method

**Zorigt Gombosuren, Khenmedekh Lochin, Aldarmaa Chuluunbaatar**

Department of Physics, Mongolian Science and Technology University, Ulaanbaatar, Mongolia

E-mail: zorigt@must.edu.mn

**Abstract**

Nonrelativistic collision of proton and antiproton with hydrogen atom described by solving time-dependent Schrodinger equation numerically. Coulomb wave function discrete variable method (CWDVR) had been used to calculate electron wave function evolution, while projectile defined classically, moving along the straight line trajectories with constant velocity. The ionization amplitude calculated by projection of the wave function into continuum wave function of the hydrogen electron. The differential cross sections calculated depending on projectile impact energy, scattering angle and electron ejection energy and angles. Our results in good agreement with the relativistic calculation results.

## 1. Introduction

Theoretical study of charged particle-atomic collision is established since early 1920. Modern experimental development technique is enabling to measure differential cross sections (DCS) of atomic collision.

The perturbative calculations of triply differential cross section (TDCS) for ionization in antiproton-hydrogen collision have been performed in Refs. S. Jones *et al* (2002) [1], A. B. Voitkiv and J. Ullrich (2003) [2].

Recently these DCS's have also been studied by several non perturbative methods. A. Igarashi *et al* (2000) [3] developed a similar approach to the one-center close-coupling (CC) calculation with larger basis sets. Xiao-Min Tong *et al* (2001) [4], solved ionization total cross section of hydrogen atom impact antiproton using general pseudospectral method (GPSM). McGovern *et al*. (2009 [5], 2010 [6]) developed a method for extracting the TDCS from an impact-parameter treatment of the collision within a coupled pseudostate (CP) formalism. Abdurakhmanov *et al*. (2011 [7]) worked out the fully quantal time-independent convergent-close-coupling (QM-CCC) approach to differential ionization studies in ion-atom collisions. Ciappina *et al*. (2013) [8] applied the time-dependent close-coupling (TDCC) technique to investigate the role of the nucleus-nucleus interaction in the TDCS. Recently, Abdurakhmanov *et al*. (2016) [9] used the semiclassical wave-packet convergent-close-coupling (WP-CCC) method to examine the TDCS. Afterwards, A. I. Bondarev *et al* (2017) [10] developed new relativistic method based on the Dirac equation for calculating TDCS's for ionization in ion-atom.

Peng. Liang-You and Starace. Anthony F, 2006 [11] are successfully applied Coulumb wave function discrete variable representation (CWDVR ) method (similar to GPSM) for laser atomic interaction.

We used a semiclassical impact parameter representation, which is mathemathically equivalent with the fully quantal momentum-transfer representation [7]. In this paper we introduce implementation of CWDVR method in antiproton-hydrogen atom collision problem first time. Combining this CWDVR method with impact parameter method, which allows computing to use ordinary personal computer.

**2. Theory**

Hydrogen-antiproton collision process is calculated by time-dependent Schrödinger equation (TDSE) expressed as follow (1).

By using impact parameter method when projectile moves along a straight-line trajectory, Hydrogen-antiproton collision problem transfers to problem of hydrogen atom in time-dependent electric field.

TDSE is expressed as follows in atomic unit system. (Here, atomic units $\hbar = 1 a.u, e = 1 a.u, m_e = 1 a.u$)

$$i\frac{\partial}{\partial t}\Psi(\vec{r},t) = [\hat{H}_0 + \hat{V}]\Psi(\vec{r},t) \quad (1)$$

Here $\Psi(\vec{r},t)$- electronic wave function, $\hat{H}_0$- Hamiltonian of hydrogen atom, $\hat{V}(\vec{r},t)$- external field.

$$\hat{V}(\vec{r},t) = \frac{-Z}{|\vec{R}(b,0,vt)-\vec{r}|} \quad (2)$$

Here $b$- impact parameter, $v$ - velocity of projectile, $\vec{r}$- electron radius vector, $\vec{R}$ projectile radius vector, t-time and if there is antiproton Z=$-1$. See figure 1 Calculating a time propagation using expression as follows.

The propagation of the wave function can be performed using second-order split-operator method.[4, 12]

$$\Psi(\vec{r},t+\Delta t) \cong exp\left(\frac{-i\hat{H}_0 \Delta t}{2}\right) \times exp\left(-i\hat{V}\left(\vec{r},t+\frac{\Delta t}{2}\right)\Delta t\right) \times$$
$$\times exp\left(\frac{-i\hat{H}_0 \Delta t}{2}\right)\Psi(\vec{r},t) + O(\Delta t^3)$$
$$(3)$$

Equation is expressed as follows in spherical coordinate system.

$$\Psi(\vec{r},t) = \sum_{l,m} R_{l,m}(r,t) Y_{l,m}(\varphi,\theta) \quad (4)$$

Here $Y_{l,m}(\varphi,\theta)$ Spherical harmonic, $R_{l,m}(r,t)$ time-dependent radial function. $H_0^l$-Hamiltonian corresponding to l is expressed as follow.

$$H_0^l = -\frac{1}{2}\frac{d^2}{dr^2} + \frac{l(l+1)}{2r^2} - \frac{1}{r} \quad (5)$$

As well known, the Hydrogen atom Hamiltionian $H_0^l$ has infinite number of discrete and continious spectrum, that are the main difficulty to use them for numerical calculations. One of solutions for this difficulty is pseudospectral method. We used CWDVR method in this problem [11].

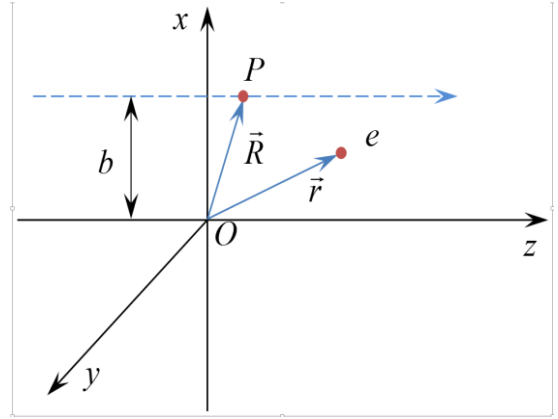

**Figure 1.** Kinematic scheme of antiproton hydrogen atom.

Here P is projectile or p- is antiproton, e- is electron. Coulomb differential equition is expressed as follow.

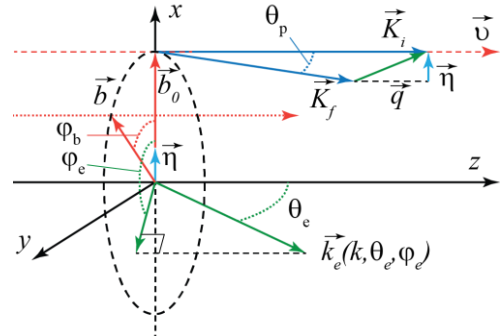

**Figure. 2**. Kinematic scheme of antiproton hydrogen atom. Antiproton is moving along z axis. $\vec{K}_i, \vec{K}_f$ – are initial and final momentum of antiproton, $\vec{k}_e$-is electron's momentum, $\vec{\eta}$ is (perpendicular to $\vec{v}$) component of the projectile momentum transfer $\vec{q}$

$$\left[\frac{d^2}{dr^2} + \frac{2Z}{r} + k^2\right]v(r) = 0 \quad (6)$$

Where k –wave number. Has the regular solution is $v(r) = F_0\left(-\frac{Z}{k}, kr\right)$. Where $F_0$ regular Coulomb wave function with parameters Z and k. The radial grid points which are the roots of the $v(r)$ and it can control the parameters.

For this grid from Eq. (5), $H_0^l$ Hamiltonian eigenvalues and eigenvectors will be defined [14].

$$H_0^l \chi_i^l = \varepsilon_i^l \chi_i^l \qquad i = 1 \dots N \quad (7)$$

Here $\chi_i^l$ is pseudospectral base corresponding to the quantum number l. N is size of base, i is spectral number.

Expanding radial function in pseudospectral base.

$$R_{l,m}(r,t) = \sum_{i=1}^{N} g_{l,m,i}(t) \cdot \chi_i^l(r) \quad (8)$$

Substituting Eq 1.50 into Eq 1.9 and defining exponential opeartor $H_0^l$.

$$exp\left(\frac{-i\widehat{H}_0 \Delta t}{2}\right) \Psi(\vec{r},t) = \sum_{l,m} \sum_{i=1}^{N} \exp(\frac{-i\Delta t}{2} g_{l,m,i}(t) \cdot \varepsilon_i^l) \chi_i^l(r) Y_{l,m}(\varphi,\theta) \quad (9)$$

We applied exponential matrix operation of Wolfram Matematica software for the computation of time propagation.

*2.1 Ionization differential cross sections*

We defeined the transition amplitude, the wave function $\Psi_k^{(-)}$ at the asymptotic time is projected on the electron wave function $\Psi(t)$.

$$T(\varepsilon, \theta_e, \varphi_e, b, \varphi_b) = \langle \Psi_k^{(-)} | \Psi(t) \rangle \quad (10)$$

Here $\varepsilon$- ejection energy or $\theta_e$, $\varphi_e$- direction of momentum $k = \sqrt{2\,\varepsilon}$, $b$- impact parameter, the angle $\varphi_b$.

Fully differential ionization probability expressed as follow.

$$\frac{d^3 P(b)}{d\varepsilon d\Omega_e db} = |T(\varepsilon,\theta_e,\varphi_e,b,\varphi_b)|^2 \quad (11)$$

Scheme shown in figure 2 is used for the differential probabilities in terms of the transverse (perpendicular to $\vec{v}$) component $\vec{\eta}$ of the projectile momentum transfer q rather than the impact parameter b.

Transition amplitudes in the $b$ and $\eta$ representations are related by a two-dimensional Fourier transform.

$$T(\varepsilon,\theta_e,\varphi_e,\eta,\varphi_\eta) = \frac{1}{2\pi} \int d\vec{b} e^{i\vec{\eta}\vec{b}} e^{i\delta(b)} T(\varepsilon,\theta_e,\varphi_e,b,\varphi_b) \quad (12)$$

Where $\delta(b)$ is the additional phase due to the projectile and target interaction.[10].

$$\delta(b) = \frac{2 \cdot z_{\bar{p}} z_p}{v} \cdot \ln(v \cdot b) \quad (13)$$

Antiproton and hydrogen atom are corresponding to $z_{\bar{p}} = -1$ and $z_p = 1$

The (fully) triply differential cross section (TDCS) may be expressed as follow.

$$\frac{d^3\sigma}{d\varepsilon d\Omega_e d\Omega_P} = K_i K_f |T(\varepsilon,\theta_e,\varphi_e,\eta,\varphi_\eta)|^2 \quad (14)$$

Therefore integrating the TDCS over corresponding variables, it can obtain various doubly differential cross sections (DDCS).

Defining DDCS which is dependent on electron ejection energy and direction of the momentum. The DDCS is independent on the antiproton scattering angle. Beacuase DDCS is obtained by integrating TDCS scattering angle.

$$\frac{d^2\sigma}{d\varepsilon d\Omega_e} = \int \frac{d^3\sigma}{d\varepsilon d\Omega_e d\Omega_P} d\Omega_P \quad (15)$$

According to the Eq. (15), the integral obtained from scattering angle is equivalent to the integral obtained by an impact parameter.

$$\frac{d^2\sigma}{d\varepsilon d\Omega_e} = \int \frac{\partial^3 P(\varepsilon,\Omega_e)}{\partial \varepsilon d\Omega_e \partial b} d\vec{b} \quad (16)$$

Define a DDCS once again when TDCS is integrating into Ejection angle and momentum transfer. The DDCS is dependet on transfer momentum $\eta$ and ejection energy.

Antiproton is moving along z axis. $\vec{K}_i, \vec{K}_f$ – are initial and final momentum of antiproton, $\vec{k}_e$-is electron's momentum, $\vec{\eta}$ is (perpendicular to $\vec{v}$) component of the projectile momentum transfer $\vec{q}$

$$\frac{d^2\sigma}{d\varepsilon d\eta} = \eta \int d\phi_\eta \int \frac{d^3\sigma}{d\varepsilon d\Omega_e d\eta} d\Omega_e \quad (17)$$

If integrating the DDCS over corresponding variables, it can obtain various singly differential cross sections (SDCS). Defining SDCS dependent on ejection angle as expressed follow.

$$\frac{d\sigma}{d\Omega_e} = \int \frac{d^2\sigma}{d\varepsilon d\Omega_e} d\varepsilon \quad (18)$$

Defining another SDCS which is dependent on ejection energy.

$$\frac{d\sigma}{d\varepsilon} = \int \frac{d^2\sigma}{d\varepsilon d\Omega_e} d\Omega_e \quad (19)$$

**3. Results**

*3.1 Details and TDCS*

In this case we chosen parameters of continuum wave functions are wave number is $k=2$ and nuclear charge is $z= 120$. Then we choose angular momentum number $l$ is 0 to 5 and radial node number is 600 respectively. Here radial maximum value $r$ was 793.3.

In Eq. (1) we solved z-component of projectile position from -80 to 560 with $\Delta z = 0.32$ step.

We chosen 225 different values of impact parameters are seleted in an interval from 0.001 a.u. to 100 a.u. The Simpson's quadrature method is used to evaluate the integral in Eq. (12) for transforming amplitudes from the impact

Journal XX (XXXX) XXXXXX | Author *et al*

parameter representation to the momentum transfer representation.

Figure 3 shows TDCS on scaterring plane calculated for antiproton energy of 200 keV, scattering angle of 0.2 mrad in the case when ejected electron's energy of 10 eV.

The polar angle $\theta_e$ of the ejected electron runs relative to direction of the momentum transfer. The direction of momentum transfer and First Born approximation maximum are correspondent to $\theta_q \approx 59.46°$.

Figure 3 shows TDCS calculation result using CWDVR is compared to results obtained in I. B. Abdurakhmanov *et al* QM-CCC [7], and A. I. Bondarev *et al* Relativistic-CC [10].

Furthermore, it is important to mention that our results are in good agreement with the results by solving the Dirac equation by A.I. Bondarev *et al*. [10]. But binary peak of QM-CCC [7] results higher than our CWDVR results. However when ejection energy is becoming lower eighter discrepancy is becoming lower [15].

In another case figure 3 shows TDCS on scaterring plane calculated for antiproton energy of 500 keV, scattering angle of 0.024 mrad in the case when ejected electron's energy of 5 eV.

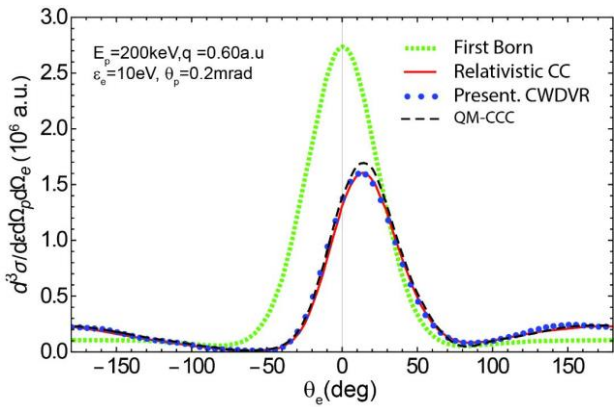

**Figure 3**. TDCS for antiproton impact ionization of hydrogen at 200 keV in the scattering plane. Results of Relativistic-CC[10] and QM-CCC [7].

When antiproton energy is 500keV present results are in good agreement with the Relativistic-CC [10] results and QM-CCC [7] results.

Figures 3 and 4 shown TDCS in scattering plane. Now figure 5 shows three dimensional TDCS, cutting a 3D surface with scattering plane. Ionization TDCS are obtained by the Eq. (14). As shown in figure 5, green surface corresponds to 4 eV ejection energy another surface corresponds to 10eV ejection energy.

Triply (fully) differential cross section is dependent on projectile scattering angle, ejection energy and ejection angle. Now we consider DDCS that dependent on two variable.

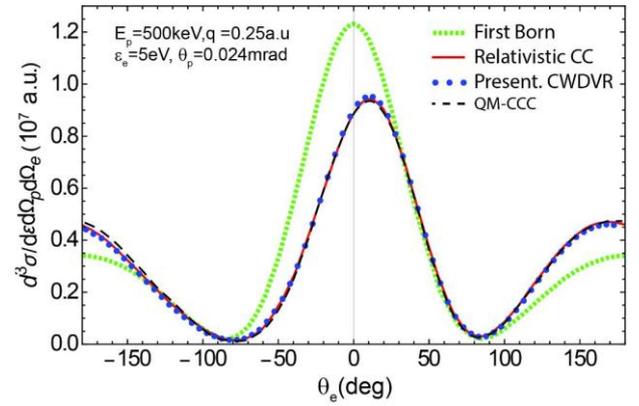

**Figure 4.** TDCS for antiproton impact ionization of hydrogen at 500 keV in the scattering plane. Results of Relativistic-CC[10] and QM-CCC [7].

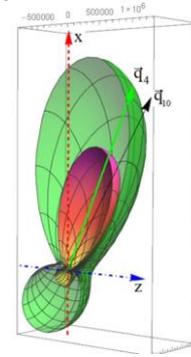

**Figure 5.** Angular distributions for ionization triply differential cross sections in three-dimensional space.

Firstly we interest in DDCS that dependent on ejection energy and ejection angle. Figure 6 shows DDCS within a range up to 10eV ejection energy calculated according to the expression (16) in present result.

*3.2 DDCS*

In our observation of present results are similar with the results of A. I. Bondarev *et al* [13] for surface shape and value. But then present results are similar with the results of I. B. Abdurakhmanov [7] for surface shape and value when ejection energy is higher than 0.1 eV.

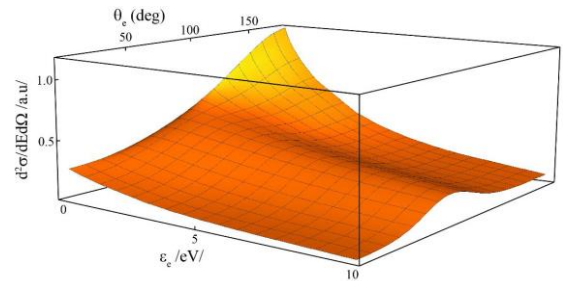

**Figure 6.** DDCS dependent on ejection energy and angle. Antiproton energy 200 keV.

As shown in figure 7, illustrates DDCS in wide range of ejection energy.

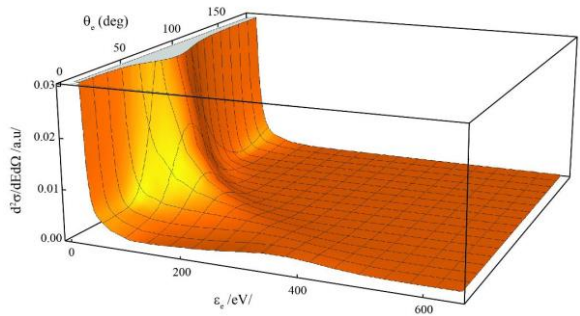

**Figure 7.** a) DDCS dependent on ejection energy and angle. When antiproton energy 200 keV.

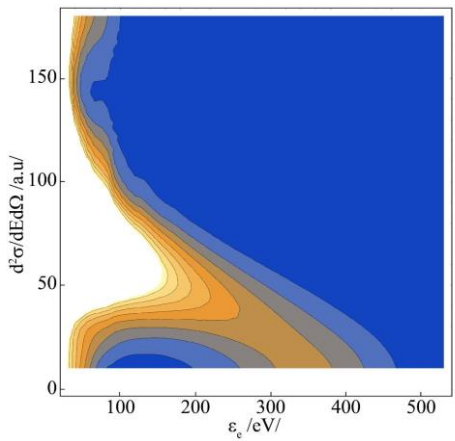

**Figure 7.** b) Contour plot of DDCS. When antiproton energy is 200 keV.

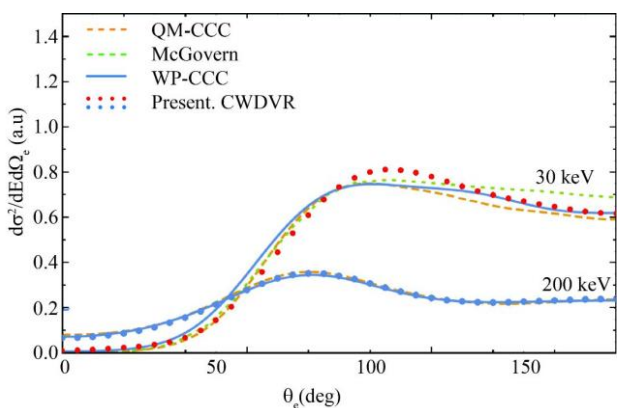

**Figure 8.** DDCS dependent on ejection energy and angle. Antiproton energy is 30 keV, 200 keV and ejection energy 5 eV. Results of McGovern [5], WP-CCC [9] and QM-CCC [7].

When ejection energy increases DDCS maximum decreases and corresponding angle θ is shifting to zero (See Figure 7). DDCS will be vanished when ejection energy value is at around 400eV. Figure 7 shows that our selected 8 wave number of continuum wave function to obtain DDCS result.

We calculated DDCS dependent on ejection energy and angle when antiproton energy is 30 keV and 200 keV. In this case ejection energy is 5 eV. Present result comparison to other theoretical results are shown in figure 8. As shown in figure 8 Present CWDVR calculation results are in a good agreement with M. McGovern *et al* (CP) [5] and I. B. Abdurakhmanov *et al* (WP-CCC) [9], (QM-CCC) [7] when antiproton energy is 200 keV. However in case of antiproton energy at 30 keV, results between M. McGovern *et al* (CP) [5] and I. B. Abdurakhmanov *et al* (WP-CCC [10], QM-CCC [7]) have discrepancy and also there is a little bit discrepancy in our calculated CWDVR results at more than $90^0$.

DDCS dependent on ejection energy and angle is calculated when antiproton energy is 500 keV, present result is coincident with M.McGovern *et al* [5].

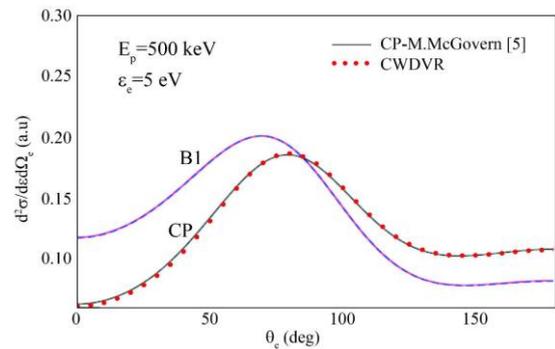

**Figure 9**. SDCS dependent on ejection angle. Results of McGovern [5] and present CWDVR.

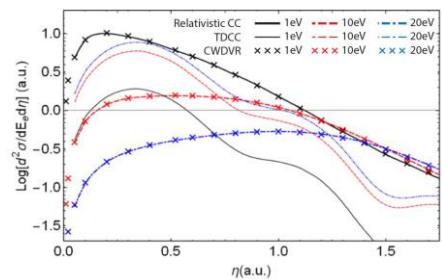

**Figure 10**. DDCS dependent on ejection energy and momentum transfer. Results of Relativistic CC [10], TDCC [8] and present CWDVR.

DDCS dependent on ejection energy and momentum transfer is calculated by Eq. (17) when antiproton energy is 200 keV and ejection energy is 1eV, 10eV, 20eV.

As shown in figure 10, the present results have good agreement with A. I. Bondarev *et al* (Relativistic CC) [10].

*3.3 SDCS*

In this case we consider in SDCS that dependent on only one variable. Firstly consider SDCS that independent on scattering angle of antiproton and ejection energy.

SDCS dependent on ejection angle is calculated by Eq. (18)

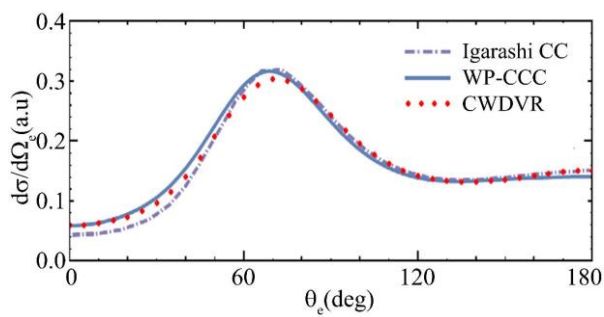

**Figure 11**. SDCS dependent on ejection angle. When antiproton energy is 200keV. Results of Igarashi [3], WP-CCC [9] and present CWDVR.

when antiproton energy is 200 keV. In order to integrate the DDSC there is a need of wide range of ejection energy. In this case we changed wave number value into 8 of continuum wave function.

We compared our calculated result of SDCS dependent on ejection angle to A. Igarashi *et al* (CC) [3], I. B. Abdurakhmanov *et al* (WP-CCC) [9] as shown in figure 11.

SDCS dependent on ejection energy is calculated by Eq. (19) when antiproton energy is 30 keV, 100 keV and 200 keV as shown in figure 12.

The Simpson's quadrature method is used to integrate by the solid angle in Eq.(19). We used 121 different values of $\varphi_e$ and 73 different value of $\theta_e$.

The SDCS decreases when ejection energy increases. But when projectile energy increases SDCS decreases. Because it is depending on interaction time of collision.

Our results are in a good agreement with M.McGovern *et al* [5, 6] and I. B. Abdurakhmanov *et al* QM-CCC [7], WP-CCC [9].

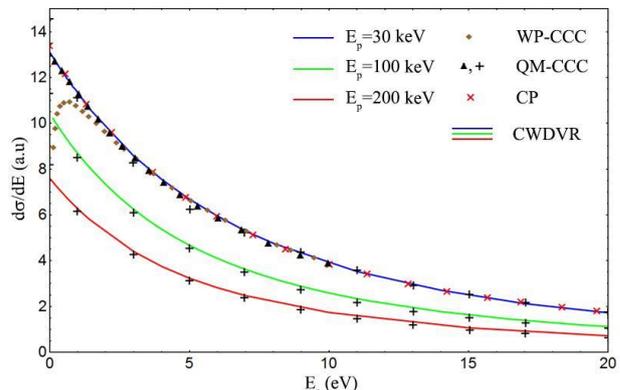

**Figure 12**. SDCS dependent on ejection energy. Results of QM-

## Conclusion

Ionization differential cross sections of antiproton impact hydrogen atom is calculated with CWDVR method by directly solving the TDSE. Present results of triply, doubly and singly cross sections have good agreement with some of non-perturbative method results such as relativistic-CC.

From the analysis of the DDCS (which depends on electron ejection energy and angle) we that conclude that the maximum of the DDCS shifts from the direction of antiproton incident at low ejection energy to the opposite direction at high ejection energy. This is the effect due to the post collision interaction between the projectile and ejected electron. Also we observed the shift of the maximum of the DDCS($\frac{d^2\sigma}{d\varepsilon d\eta}$) to the higher value of the transferred momentum $\eta$ with the increase of the electron ejection energy. We explain this shift as the effect of the momentum conservation law.